\def\be{\begin{equation}}
\def\ee{\end{equation}}
\def\bea{\begin{eqnarray}}
\def\eea{\end{eqnarray}}
\def\a{\alpha}

\def\d{\delta}

\def\p{\phi}

\def\t{\theta}
\def\s{\sigma}

\def\x{\psi}

\documentclass[12pt]{article}

\begin{document}

 \title{The random cluster model \\
and \\
new summation and integration identities}
 \vskip 1cm
\author{L. C. Chen, Institute of Mathematics, Academia Sinica \\ Taipei 11529, Taiwan\\
and \\ F. Y. Wu, Department of Physics \\
Northeastern University, Boston, Massachusetts 02115, U.S.A.}
\date{}
\maketitle
 
\begin{abstract}
We explicitly evaluate  the free energy of the random
cluster model at its critical point for $0<q<4$
using an exact result due to Baxter, Temperley and Ashley.  
It is found that the resulting expression assumes a form which
 depends on whether  $\pi /2\cos ^{-1} (\sqrt q / 2)$ 
is a rational number, and if it is a rational number whether the denominator 
is an odd integer.  Our consideration lealds to   
  new summation identities and, for $q=2$, a closed-form evaluation of the integral
 \bea
&&\frac 1 {4\pi^2} \int_0^{2\pi}d\theta \int_0^{2\pi}d\phi 
  \ln \big[A+B+C-A\cos \theta - B\cos \phi -C \cos({\theta+\phi} ) \big] \nonumber \\
&&\quad\quad\quad  = - \ln (2S)  
+ (2 /\pi) \big[ {\rm Ti}_2 (AS) +{\rm Ti}_2 (BS) +{\rm Ti}_2 (CS)  \big], \nonumber
\eea
where $A,B,C\geq 0$ and $S=1/\sqrt{AB+BC+CA}$.
 \end{abstract}

\newpage
\section{Introduction}
The consideration of exact solutions of lattice models in statistical mechanics is a
fertile field in both physics and mathematics.  In physics it is well-known 
 that (see, for example, \cite{baxter}) exact solutions of lattice models lead 
to the finding of different classes of critical phenomena.
In mathematics exact solutions in statistical mechanics 
often lead to the discovery of new mathematical identities.
A prominent example is the solution
of the hard hexagon model by Baxter \cite{BaxterHH}
leads to the discovery of new Rogers-Ramanujan summation identities in combinatorics \cite{BaxterRam}.
As another example consider the integral
\bea
I(A,B,C)&=&\frac 1 {4\pi^2} \int_0^{2\pi}d\theta \int_0^{2\pi}d\phi \ln \Big[A+B+C \nonumber \\
&& \hskip 1.5cm  -A\cos \theta - B\cos \phi -C \cos({\theta+\phi} ) \Big] \label{IABC}
 \eea 
which arises often in lattice statistical studies.
For $A=B=2, C=0$ it is well-known \cite{kas,fisher} that
the integral gives the per-site entropy of dimers on a square lattice 
and the integral can be directly evaluated \cite{kas,fisher} to yield
\bea
I(2,2,0) = \frac 4 \pi \bigg(1 - \frac 1 {3^{2}}
 + \frac 1 {5^{2}} - \frac 1 {7^{2}} - \cdots \bigg)\ . \label{220}
\eea
 For $A=B=C=2$ the integral gives the per-site entropy of spanning trees on 
the triangular lattice, and the exact solution of the latter problem \cite{wu,glasserwu} 
   yields the closed form evaluation
\bea
 I(2,2,2)= \frac 6 \pi {\rm Ti}_2\bigg(\frac 1 {\sqrt 3} \bigg) + \frac 1 2 \ln 3 ,\label{222}
\eea
where ${\rm Ti}_2$ is the inverse tangent integral function  \cite{lewin} defined by
\begin{eqnarray}
{\rm Ti}_2(a) &=& \int_0^a\frac{\tan^{-1}t}{t}\;dt \nonumber \\
&=&  a - \frac {a^3} {3^2} +\frac {a^5} {5^2} - \frac {a^7} {7^2} +\cdots . \label{arctan}
 \end{eqnarray}
Closed-form evaluation of $I(A,B,C)$ for general $A,B,C$ is not known.
 
\medskip
In this paper we report a closed-form evaluation of $I(A,B,C)$, 
by using a result due to
  Baxter, Temperley and Ashley \cite{bta} on the random cluster model  \cite{kf,fk}.
 We also show that the explicit expression of its critical free energy 
depends crucially on the value of $q$. This implies that the free energy
of the random cluster model, if solved, would also share this property.
  For convenience of references, we first state the integration result:
 
\medskip
{\it For arbitrary $A,B,C \geq 0$ and $S = 1/ \sqrt {AB+BC+CA}$, we have}
\bea
 I(A,B,C) = - \ln (2S)  
+ \frac 2 \pi \bigg[ {\rm Ti}_2 (AS) +{\rm Ti}_2 (BS) +{\rm Ti}_2 (CS)  \bigg]. \label{q2}
\eea

\section{The random cluster model}
Consider a $q$-state Potts model on the triangular lattice of $N_s$ sites with anisotropic (reduced) interactions
$K_1,K_2,K_3$ along the three principal lattice directions. Its  partition function is
\be
Z^P_{N_s}(K_1,K_2,K_3) = \sum _{\s_i=1}^q \prod_{E} {\rm exp}[K_\a \d_{\rm Kr} (\s_i,\s_j)] \label{part}
\ee
where the summation is over all $q^{N_s}$ spin states $\s_i$, $i=1,2,...,q$, and the product is over nearest-neighbor set $E$
connecting sites $i$ and $j$,
$\a = 1,2,3$, and $\d_{Kr}$ is the Kronecker delta function. 
We are interested in the ferromagnetic regime 
$K_\a \geq 0$.

\medskip
The partition function  can be rewritten \cite{bta} in a graph expansion as 
\be
Z^{ RC}_{N_s}(K_1,K_2,K_3) = \sum_G q^{c(G)} (e^{K_1}-1)^{\ell_1}(e^{K_2}-1)^{\ell_2}(e^{K_3}-1)^{\ell_3} 
\label{part1}
\ee
where the summation is over all subgraphs $G$ of the triangular lattice, 
$c(G)$ is the number of connected clusters in $G$ including isolated points,
and $\ell_\a$ is the number of lines in $G$ in the direction $\a=1,2,3$.
In this form the partition function  describes that of a random cluster model
\cite{kf,fk} for which $q$ can be continuous.
 One is interested in   the per-site ``free energy" of the random
cluster model
\be
f^{ RC} = \lim_{{N_s}\to \infty} N_s^{-1} \ln Z^{ RC}_{N_s}(K_1,K_2,K_3).\label{RCfree}
\ee

The triangular random cluster model is known 
\cite{bta,wu1} to be critical at 
\be
\sqrt q\ x_1x_2x_3 +x_1x_2+x_2x_3+x_3x_1 =1 \label{criticalpoint}
\ee
where 
\be
x_\a = (e^{K_\a}-1)/\sqrt q \geq 0. \label{x}
\ee
For $0<q<4$, define $\p(q)$ and $v_\a(q)$, $\a=1,2,3$, by
\bea 
\cos \p(q) &=& \sqrt q/2\ , \hskip 2.7cm 0< \p <\pi/2 \nonumber \\
x_\a &=& \sin (\p-v_\a) / \sin v_\a\ ,  \quad 0 < v_\a < \p. \label{vrange}
\eea
For integral values of $q$, for example, we have 
\bea
q&=&1,  \hskip 1cm 0<v_a < \p(1) =\pi /3 \nonumber \\
q&=&2, \hskip 1cm 0<v_a < \p(2) = \pi /4 \nonumber \\
q&=&3,  \hskip 1cm 0<v_a < \p(3) =\pi /6 .\nonumber 
 \label{qphi}
 \eea
In terms of these variables, we have
 \bea
e^{K_\a} &=& 1 + \frac 1 2 \Big[ \sqrt{q(4-q)} \cot v_a -q\Big], \quad 0<q<4 \label{v}
\eea
and the critical point (\ref{criticalpoint}) can be written
as
\be
v_1+v_2+v_3 = 2\p(q). \label{criticalpoint1}
\ee

\section{The critical free energy}
 Baxter, Temperley and Ashley \cite{bta} considered the random cluster model 
(\ref{part1}) on the triangular lattice
and obtained, among other things, its
free energy at the critical point (\ref{criticalpoint}).  For $0<q<4$ they obtained
\be
f^{ RC}_{critical}=
\frac 1 2 \ln q + \x(\p,v_1)+\x(\p,v_2)+\x(\p,v_3) \label{RCfree}
\ee 
where
\be
\x (\p,v) = \frac 1 2 \int_{-\infty}^\infty
\frac { \sinh (\pi-\phi) x \ \sinh 2(\phi-v)x }
      { x \sinh \pi x\ \cosh\phi x } dx, \quad 0<q<4. \label{xintegral}
\ee
   Here, we investigate implications of 
  (\ref{RCfree}) and (\ref{xintegral}).
 
\medskip
The  integral (\ref{xintegral}) can be evaluated by contour integration.
Let  $C$ be the contour consisting of the real axis and the half circle at infinity encircling the upper half
  $z$ plane.
In the variable range $0<\p<\pi/2, 0<v < \p$, we can neglect the contribution from
the half-infinite circle and rewrite (\ref{xintegral}) as 
 \begin{eqnarray}
\psi (\p,v) = \frac 1 2 \int_{C}\frac { \sinh (\pi - \p)  z\  \sinh 2(\p-v)z }
      { z \sinh \pi z \ \cosh\p z } dz, \label{contour}
\end{eqnarray}
where the integration is alone the contour $C$.
 
\medskip
The contour integral can be carried out by using the residue theory.
The integrand in (\ref{contour}) has poles at
\begin{eqnarray}
\sinh (\pi z_1)=0&& \qquad \mbox{ or }\quad z_1=ni, \hskip 2.5cm n=1,2,..., \nonumber \\
\cosh (\p z_2)=0&&\qquad \mbox{ or }\quad  {z_2}  =\frac \pi {2\p} \big(2m+ 1 \big)  i,
\quad\quad m=0,1,2,.... \nonumber 
\end{eqnarray}
The evaluation of the integral thus depends on whether $z_1$ and $z_2$ overlap.
Overlapping of $z_1$ and $z_2$ occurs
 if $\pi/2 \p$ is a rational number with the denominator an odd integer ($M$ and $N$ have no common
integer factors):
\be
\frac \pi {2\p} = \frac M N, \quad M = 1,2,3,\cdots, \ N= 1,3,5, \cdots, \ 
N< M .\label{overlap}
\ee

Then there are two cases to consider.
 
\medskip
{\it Case 1}. 
There is no overlap between $z_1$ and $z_2$, namely, $\pi /{2\p}$ is either irrational
or is of the form of (\ref{overlap}) but with $N$ even. This includes the $q=1$ Potts model
($M=3,\ N=2$).  Then both $z_1, z_2$ are simple poles, 
and the residues can be computed straightforwardly. 
One obtains
\bea
\x(\p,v) &=& \sum_{n=1}^\infty \frac 1 n
\tan (n\p) \sin 2n(\p-v) \nonumber \\
&& \quad + \sum_{m=0}^\infty \frac 2 {2m+1} \cot \Big[\Big( m+\frac 1 2 \Big) \frac {\pi^2} \p\Big]
\ \sin \Big[ (2m+1) \frac {v\pi} \p\Big], \label{noteven}
\eea
where the two terms come from the evaluation of residues at $z_1$ and $z_2$, respectively.

\medskip
For later use the expression (\ref{noteven}) can be further simplified by introducing in the first term the
identity
\bea
\tan x \sin 2(x-y) = \cos 2y - \cos 2(x-y) +\tan x \sin 2y \nonumber
\eea
with $x= n \p,\ y = nv$.
This leads to 
\bea
\x(\p,v) &=& \ln \Bigg[\frac {\sin(\p-v)}{\sin v}\Bigg] + \sum_{n=1}^\infty \frac 1 n
\tan (n\p) \sin 2nv \nonumber \\
&& \quad + \sum_{n=0}^\infty \frac 2 {2n+1} \cot \Big[\Big( n+\frac 1 2 \Big) \frac {\pi^2} \p\Big]
\ \sin \Big[ (2n+1) \frac {v\pi} \p\Big] \label{noteven1}
\eea
after making use of  the summation identity (1.441.2 of \cite{gr})
\bea
\sum_{n=1}^\infty \frac {\cos n x} n = - \ln \Big[ 2 \sin \Big( \frac x 2 \Big) \Big],\quad 0<x<2\pi.\nonumber
\eea

\medskip
{\it Case 2}. There is overlapping between $z_1$ and $z_2$. 
This occurs when  (\ref{overlap}) holds with $N=$ odd so that when
 \be
 2m+1 = N, 3N, 5N \cdots \label{z2}
\ee
we have $z_2 = Mi,\, 3Mi,\, 5Mi, \cdots$ which coincide with $z_1$  
 where they form double poles. Note that when $N=1$ every point in $z_2$ is a double pole.
This includes the $q=2,3$ ($M=2,3,\ N=1$) Potts models.
  The integral (\ref{contour}) now consists of three terms,
$\x_2 ( {N\pi}/{2M},v) = R_{S1} + R_D +R_{S2},$
where $R_{S1}$ is the sum of residues from simple poles in $z_1$,  $R_D$ the  residues
from double poles, and $R_{S2}$ the  residues from simple poles in $z_2$, if any.
 
\medskip
To compute $R_{S1}$, we note that the forming of double poles
excludes
points   $ Mi,\ 3Mi,\ 5Mi, \cdots$ which divide the remaining $z_1 = ni$ into sections
$n=\{1,M-1\}, \{M+1, 3M-1\}, \{3M+1, 5M-1\}, \cdots$.
  Then  we can write 
\be
R_{S1} = R_{S1{\rm a}}+R_{S1{\rm b}}
\ee
where $R_{S1{\rm a}}$ is the sum of the first $M-1$ residues  and $R_{S1{\rm b}}$ is the sum of the rest.
From we have already computed in the first term in
(\ref{noteven}), we obtain
 \bea
R_{S1{\rm a}} &=&  \sum_{k=1}^{M-1} \tan (k\phi )\bigg[\frac{\sin [2k(\phi-v)] } k \bigg], 
\hskip1cm \p=\frac {N\pi} {2M} \nonumber \\
   R_{S1{\rm b}} &=&\sum_{k=-(M-1)}^{M-1}\ \sum_{n=1}^{\infty}\tan[(2nM
+k)\phi]\bigg[\frac{ \sin [2(2nM+k)(\phi-v)] } { 2nM+k}\bigg].\nonumber
\eea
Expanding the sine function, we can rewrite $R_{S1{\rm a}}$  as 
\bea
R_{S1{\rm a}} 
   = \sum_{k=1}^{M-1}\tan (k\phi)\biggl[\sin (2k\phi)
\int_{2v}^{{\pi}/{2k}}\sin kx dx-\cos (2k\phi)\int_0^{2v}\cos kx  dx\bigg].
\label{s11}
\eea
To evaluate $R_{S1{\rm b}}$ we use $4M\p = 2N\pi$ and 
 proceed similarly as in (\ref{s11}), and obtain
\bea
R_{S1{\rm b}} &=&\sum_{k=-(M-1)}^{M-1}\ \tan(k\phi) \sum_{n=1}^{\infty} 
\bigg[\sin(2k\phi)\int_{2v}^{{\pi}/{2|k|}}\sin [(2nM+k)x]dx \nonumber \\
&& \hskip 2cm -\cos (2k\phi)\int_0^{2v}\cos [(2nM+k)x]  dx \bigg],\quad \p=\frac{N\pi}{2M}\nonumber 
\eea
where one verifies that the integrated contributions   at the upper limit $\pi/2|k|$ 
cancel out exactly after combining terms from $k$ and $-k$. Here the $k=0$ term
vanishes due to the $\tan (k\p)$ factor.

\medskip
Expanding out the sine and cosine functions as before,
terms odd in $k$ in the summand are canceled and terms even in $k$ give the same contribution for 
$k>0$ and $k<0$. Thus we have 
  \bea
R_{S1{\rm b}}&=&2\sum_{k=1}^{M-1}\tan(k\phi) \sum_{n=1}^{\infty} 
\bigg[\sin(2k\phi)\int_{2v}^{{\pi}/{2k}}\sin (2nMx)\cos (kx) dx \nonumber \\
&& \hskip 1cm +\cos (2k\phi)\int_0^{2v}\sin(2nMx)\sin (kx) dx\bigg].\label{s122}
 \eea
Write
\bea
\sum_{n=1}^\infty \sin (2nMx) &=& \lim_{{\cal N}\to\infty} \sum_{n=1}^{\cal N} \sin (2nMx)
\nonumber \\    &=& \lim_{{\cal N}\to\infty} \frac {\cos (Mx)-\cos [(2{\cal N}+1)Mx]}{2\sin Mx},
\label{sinesum}
\eea
 where we have used  the summation identity 1.342.1 of \cite{gr} in writing down the last step.  
Substituting (\ref{sinesum}) into (\ref{s122}), 
  terms with the factor $\cos[(2{\cal N}+1)Mx] $ 
inside an integral vanish in the ${\cal N}\to\infty$ limit.
 Thus, one obtains 
\bea
R_{S1{\rm b}} &=& \sum_{k=1}^{M-1}\tan (k\phi)\bigg[
\sin(2k\phi)\int_{2v}^{{\pi}/{2k}} {\cos (k x)\cot(Mx)} dx \nonumber \\
 && \hskip 2cm +\cos (2k\phi)\int_0^{2v}
{\sin (kx)\cot (Mx)} dx \bigg].\label{s12}
\eea
Finally, combining (\ref{s11}) and (\ref{s12}), we obtain
\bea
R_{S1}&=& \sum_{k=1}^{M-1}\tan (k\phi)\bigg[\sin(2k \phi)\int_{2v}^{ {\pi}/{2k}}
\frac {\cos(M-k)x}{\sin (Mx)}dx\nonumber \\
&& \hskip2cm  -\cos(2k \phi)\int_0^{2v}\frac {\sin(M-k)x}{\sin (Mx)}dx\bigg], 
\quad \p=\frac {N\pi} {2M}.\label{s1f}
\eea
 
Now we compute $R_D$. 
Residues from double poles at $z_2= (2m+1)Mi$ are
\bea
R_D = \pi i \sum_{m=0}^\infty \lim _{z\to z_2}
\frac d {d z} \Bigg[ (z-z_2)^2 \Bigg(
\frac { \sinh (\pi - \p)  z\  \sinh 2(\p-v)z }
      { z \sinh \pi z \ \cosh\p z }\Bigg)  \Bigg]. \label{s21}
\eea
 where $z_2=(2m+1)Mi,\ \p = N\pi/2M,\ N=$ odd.

\medskip
Define number $u$ and integer $p$ by
\bea
  Mv = p\pi/2 + u, \quad{\rm where} \quad 0< u < \pi/2, \ p=0,1,2,\cdots,\  p<N. \label{Mv}
\eea
Here for technical reasons we exclude points $u=0$. At  these points each of $R_{S1}$, $R_D$ and $R_{S2}$
diverge. But in this cse $\x(\p,v)$ can be computed by taking an appropriate large
$M,N$ limit of (\ref{noteven1}). (See discussions in Section 5 below.)
 Then using the identities
  \bea
 \lim_{z\to z_2} \frac {z-z_2} {\sinh \pi z} &=& \frac 1 \pi (-1)^M ,\hskip2.1cm
 \lim_{z\to z_2} \frac {z-z_2} {\cosh \p z} = \frac {2M} {N\pi i} (-1)^{m+(N-1)/2}  \nonumber\\
  \sinh (\pi -\p)z_2 &=& i(-1)^{M-m+(N+1)/2}, \quad \sinh 2(\p -v)z_2 = i(-1)^p\sin 2(2m+1)u \nonumber \\
 \cosh  (\pi -\p)z_2 &=& 0\,, \hskip 3.4cm \cosh  2(\p -v)z_2 =(-1)^{p+1}\cos 2(2m+1)u\nonumber 
 \eea
it is straightforward to obtain from (\ref{s21})
\be
      R_D = \frac {2(-1)^p} {MN\pi} \sum_{m=0}^{\infty}\frac {\sin[2(2m+1)u]} {(2m+1)^2}
          + \frac {4(-1)^p(\p-v)} {N\pi} \sum_{m=0}^\infty \frac {\cos [2(2m+1)u] }   {2m+1} . \label{s22}
\ee

The first summation in (\ref{s22})
can be carried out by using the summation identity 3.1 established below, and the 
second summation   carried out using 1.442.2 of \cite{gr}, namely,
\be
\sum_{k=0}^\infty \frac {\cos2(2k+1)x} {2k+1} =\frac 1 2 \ln \cot x\ , \quad 0<x<\frac \pi 2 .\label{cossum}
\ee
 This leads to the simple expression
  \be
         R_D = \frac {2(-1)^p}{MN\pi}{\rm Ti}_2(\tan u)+\frac {(-1)^p(N-p)}{MN}\ln \cot u. \label{s2f}
\ee
 
\medskip
{\it Identity 3.1.}
\be  
\sum_{k=0}^\infty \frac {\sin 2(2k+1) x} {(2k+1)^2}
= {\rm Ti}_2(\tan x) + x \ln \cot x, \quad 0<x<\frac \pi 2. \label{identity2}
 \ee
 
\medskip
{\it Proof.} We have 
 \bea
\sum_{k=0}^{\infty}\frac {\sin 2(2k+1)x}{(2k+1)^2}
 &&=\int_0^{2x}\sum_{k=0}^{\infty}\frac {\cos (2k+1)y}{2k+1}dy\nonumber\\
&&=\frac 1 2 \int_0^{2x}\ln\Big(\cot\frac {y}{2}\Big)dy\nonumber\\
 &&=-\int_0^{\tan x}\frac {\ln t}{1+t^2}\ dt\ , \quad 0<x<\frac \pi 2 ,
\eea
where we have made use of (\ref{cossum}) and the change of variable $t=\tan (y/2)$.
This leads to (\ref{identity2}) after using the identity (4.29) of \cite{lewin}, namely,
\be
 {\rm Ti}_2(\tan x)=x \ln (\tan x) -\int_0^{\tan x}\frac {\ln t}{1+t^2}\ dt, \quad x<\frac \pi 2.
\ee
 Q.E.D.

  \medskip
To compute $R_{S2}$, the sum of residues of simple poles in $z_2$,
we need to exclude from $z_2=[(2m+1)M/N]i$ the double poles $Mi,\
3Mi,\ 5Mi, \cdots$. As a result, the remaining simple poles in $z_2$ are divided into sections
 $m = \{0, (N-3)/2\}, \{(N+1)/2, (3N-3)/2\}, \{(3N+1)/2, (5N-3)/2 \}, \cdots$.
The situation is similar to that of $R_{S1}$, and $R_{S2}$ can be similarly computed.
   Omitting the details we 
arrive at the expression
\bea
R_{S2}  =-\frac {2M} N \sum_{k=1}^{{(N-1)}/2} \cot \bigg(\frac {2kM\pi}{N} \bigg)\int_0^{2v}
\frac {\sin  (2kMx/N)}{ \sin (Mx) }dx,
\quad N=1,3,5,\cdots. 
\label{R3}
\eea
Note that $R_{S2}=0$ when $N=1$, as every point in $z_2$ is a double pole.

Finally, combining the above results, we obtain the expression
\bea
\x \bigg( \frac {N\pi}{2M}, v\bigg)=R_{S1}+R_D  + R_{S2},\quad M=1,2,3,\cdots \quad N =1,3,5,\cdots
\nonumber \\
\label{psi33}
\eea
where $R_{S1}$ is given in (\ref{s1f}), $R_D$ by (\ref{s2f}), and
$R_{S2}$ by (\ref{R3}).

\section{Special cases}
In this section we consider the special cases of $q=1,2,3$ Potts models.

\medskip
For $q=1$, we have $\p = \pi/3, \, M=3,\, N=2$, so
we use (\ref{noteven1}). For $\p=\pi/3$, we have  $\cot \big[( n+ 1/ 2 ) {\pi^2} /\p\big] =0$
and the third term in (\ref{noteven1}) vanishes identically.
It is then a simple matter to use the summation identity 4.1 we establish below to obtain
\be
\x\Big(\frac \pi 3, v\Big) = \ln \frac {\sqrt 3 \cot v +1} 2.
\ee
It follows that the free energy is
\bea
f ^{RC} = K_1+K_2+K_3\ ,\hskip 1cm q=1 \nonumber
\eea
after making use of (\ref{v}) and (\ref{RCfree}). This result agrees with the physical expectation that,
since  there is only one spin state  in the $q=1$ Potts model, the partition function is 
$Z={\rm exp} [N_s(K_1+K_2+K_3)]$.
 
\medskip
{\it Identity 4.1.}
\be  
\sum_{n=1}^\infty \frac 1 n
\tan (n\pi/3) \sin 2nv  = \ln \frac {\sqrt 3 \cot v+1}{\sqrt 3 \cot v-1},
\quad 0<v< \pi /6 . \label{identity1}
\ee

\noindent
{\it Proof.}
Since $\tan (n\pi/3) = 0 $ for $n= 3 \times {\rm integers}$, we have
 \begin{eqnarray}
\sum_{n=1}^\infty \frac 1 n
\tan (n\pi/3) \sin 2nv  &=& \sqrt 3
\sum_{n=0}^{\infty}\bigg[\frac {\sin 2(3n+1)v}{3n+1}-\frac {\sin 2(3n+2)v}{3n+2}\bigg] \nonumber\\
&=&\sqrt{3}\sum_{n=0}^{\infty}\bigg[\int_0^{2v}\Big[\cos (3n+1)x-\cos (3n+2)x\Big] dx\nonumber\\
&=& \sqrt 3 \lim_{{\cal N}\to \infty} \sum_{k=0}^{{\cal N}-1}\int_0^{2v} 
 \Big[\cos (3k+1)x-\cos (3k+2)x\Big] dx \nonumber
\label{lemma1}
\end{eqnarray}
 The summation can be carried out by using the formula (1.341.3 of \cite{gr})
\bea
\sum_{k=0}^{n-1}\cos (x+ky)=\cos\Bigg(x+\frac {n-1} 2 y\Bigg)\sin\frac {ny}{2}\csc \frac y{2}\ . 
\nonumber 
\eea
This yields
\begin{eqnarray}
\sum_{n=1}^\infty \frac 1 n
\tan (n\pi/3) \sin 2nv  &=& \sqrt 3
 \lim_{{\cal N}\to\infty}\int_0^{2v}\frac
 {\bigl(1-\cos 3{\cal N}x\bigr)\sin ( x/2)} {\sin ({3x}/{2})} dx\nonumber\\
&=&\sqrt 3\int_0^{2v}\frac {dx}{2\cos x+1}-\sqrt 3\lim_{{\cal N}\to\infty}\int_0^{2v}
\frac {\cos 3{\cal N}x}{2\cos x+1}dx. \nonumber \\
\label{lemma2} 
\end{eqnarray}
The second term in (\ref{lemma2}) vanishes and the
first term can be evaluated by using the formula (Cf. 2.551.3 of \cite{gr})
\bea
\int\frac {dx}{a+b\cos x}=\frac {1}{\sqrt{b^2-a^2} }
\ln \Bigg[\frac { \sqrt{b^2-a^2}\tan (\frac x2 )+a+b}
{ \sqrt{b^2-a^2}\tan (\frac x2 )-a-b}\Bigg],\quad a^2<b^2.\nonumber 
\eea
This leads to (\ref{identity1}). Q.E.D.

\medskip
For $q=2$ we have $\p = \pi/4, \, M=2,\ N=1$.
 By comparing our result with that of the Ising model, 
the  critical free energy  leads to the integration formula (\ref{q2})
as we shall now show.

\medskip
First, using (\ref{psi33}) we obtain
 \be
\psi\bigg(\frac \pi 4,v\bigg) = \frac 1 2 \ln[(\cot v)(\cot 2v)] +
\frac 1 \pi {\rm Ti}_2 (\tan 2v) , \quad q=2.
\ee
 Therefore from  (\ref{RCfree}) 
the critical free energy is 
\be
f^{Potts} = \frac 1 2 \ln 2 + \sum_{\a=1}^3 \bigg[\frac 1 2 
  \ln[ (\cot v_\a )(\cot 2v_a)] + \frac 1 \pi {\rm Ti}_2(\tan 2v_a) \bigg], \quad q=2 .\label{potts1}
\ee

On the other hand, the $q=2$ Potts model is completely equivalent to an Ising model.
Consider an Ising model on the same triangular lattice of $N_s$ site with anisotropic interactions
$K_\a/2$, $\a=1,2,3$. We have the equivalence
\bea
Z_{N_s}^{Ising} &=& \sum _{\s = \pm 1} \prod_E e^{(K_\a/2)\s_i\s_j} \nonumber \\
   &=& e^{-N_s(K_1+K_2+K_3)/2} Z^{Potts}_N \Big|_{q=2}
\eea
after making use of the identity $\s_i\s_j= 2\d_{\rm Kr}(\s_i,\s_j)-1$.
It follows that their critical free energies are related by
\bea
f^{Ising} &=&  f^{Potts}\Big|_{q=2}- \frac 1 2 (K_1+K_2+K_3)  \nonumber \\
&=&  f^{Potts}\Big|_{q=2}-\frac 1 2\ln\big[(\cot v_1)(\cot v_2)(\cot v_3)]  \nonumber \\
&=& \frac 1 2 \ln 2+\sum_{\a=1}^3 \bigg[ \frac 1 2 \ln (\cot 2v_\a)
+\frac 1 \pi {\rm Ti}_2 (\tan 2v_a)\bigg] 
\label{equalf}
\eea
after substituting with (\ref{potts1}) and using (\ref{v}) for $q=2$, namely,
\be
 e^{K_\a} = \cot v_\a \ , \quad {\rm or} \quad \sinh K_\a = \cot 2 v_\a .\label{Ka}
\ee

\medskip
In fact, the Ising free energy is known at all temperatures \cite{hou} to be
\bea 
f^{Ising} &=& \ln 2 + \frac 1 {8\pi^2} \int_0^{2\pi}\int_0^{2\pi}\ln \Big[ \cosh K_1\cosh K_2\cosh K_3
+\sinh K_1\sinh K_2\sinh K_3\nonumber \\
   && \quad  -\sinh K_1 \cos \theta -\sinh K_2 \cos \p -\sinh K_3\cos(\t+\p)\Big]d\t d\p. \label{Isingfe}
\eea
   It can  be verified that the critical condition  (\ref{criticalpoint})
is equivalent to
\bea
&&\cosh K_1\cosh K_2\cosh K_3
+\sinh K_1\sinh K_2\sinh K_3 \nonumber \\
&& \hskip1cm =\sinh K_1+\sinh K_2+ \sinh K_3, \label{criticalpoint2}
\eea
so  we can rewrite (\ref{Isingfe}) as
\be 
f^{Ising} = \ln 2 + \frac 1 2 I(a,b,c) \label{Isingf}
\ee
 with $a=\cot 2v_1, b=\cot 2v_2, c = \cot 2v_3$ and $I(a,b,c)$  defined in (\ref{IABC}).
It can also be verified that the critical condition $v_1 + v_2+ v_3 = \pi/2$
is the same as
\be
ab + bc+ca =1. \label{abc}
\ee
Equating (\ref{equalf}) with (\ref{Isingf}), we obtain
\be
I(a,b,c) = - \ln 2 +\ln (abc) + \frac 2 \pi \Big[{\rm Ti}_2(a^{-1})
+{\rm Ti}_2(b^{-1})+ {\rm Ti}_2(c^{-1}) \Big].
\ee

\medskip
For the  general integral  $I(A,B,C)$ 
 we can always define variables 
$a=AS, b=BS, c=CS$ with $S=1/\sqrt {AB+BC+CA}$  to satisfy (\ref{abc}). 
Thus, one has
\bea
I(A,B,C) &=& -\ln S +I(a,b,c) \nonumber \\
&=& -\ln (2S) + \ln (abc)+ \frac 2 \pi \Big[{\rm Ti}_2(a^{-1})
+{\rm Ti}_2(b^{-1})+ {\rm Ti}_2(c^{-1}) \Big] \nonumber \\
&=& -\ln (2S)+ \frac 2 \pi \Big[{\rm Ti}_2(a)
+{\rm Ti}_2(b)+ {\rm Ti}_2(c) \Big], \label{proof}
\eea
where  use has been made   of the identity (Cf. (2.6) of \cite{lewin})
\bea
{\rm Ti}_2 (y^{-1}) = {\rm Ti}_2 (y) - \frac \pi 2 \ln y, \quad y>0 .\nonumber
\eea
 The last line in (\ref{proof})  establishes  the integration formula (\ref{q2}). 
It is readily checked that  (\ref{proof}) yields
the previous known values of $I(2,2,0)$ and $I(2,2,2)$.
 
\medskip
For $q=3$ we have $\p = \pi/6, \,  M=3, \, N=1$, so we again use (\ref{psi33}). 
It is straightforward to obtain
\bea
\psi \bigg(\frac {\pi} {6},v\bigg)&=&\frac {1}{6}\ln\frac {2\cos 2v+\sqrt {3}}{2\cos 2v-\sqrt {3}}
+\frac 3 2 \ln\frac {{\sqrt 3}+1}{{\sqrt 3}-1}
-\ln\frac {{\sqrt 3}\cot 2v+1}{{\sqrt 3}\cot 2v -1}\nonumber\\
&+&\frac {1}{12}\ln \frac { 2-{\sqrt 3} }{ 2+{\sqrt 3} }
+\frac 13\ln\cot 3v+\frac 2{3\pi}{\rm Ti}_2(\tan 3v), \ \, v<\frac\pi 6. \label{pq3}
\eea
 
\section{Summary and acknowledgments}
We have evaluated explicitly  the free energy of the random
cluster model at its critical point for $0<q<4$
using an exact result  of Baxter, Temperley and Ashley.  
It is found that the critical free energy  is given by
   (\ref{noteven}) if $\pi/2\p(q)=\pi/2 \cos ^{-1} (\sqrt q / 2)$
is irrational or a rational number with an even denominator, 
and  by (\ref{psi33}) if $\pi/2 \p $
is a rational number with an odd integer in the denominator.
  Special cases of our results lead to 
  new summation identities
(\ref{identity2}) and (\ref{identity1}), and a new integration formula (\ref{q2})
for $I(A,B,C)$, which do not appear to have
previously appeared in print.

\medskip
It is instructive to see how  the critical free energy passes from (\ref{psi33})
to (\ref{noteven})  as $q$ varies and $\pi/2\p$ changes from rational to irrational. 
 Indeed, any irrational $\pi/2\p$  can be reached by taking an appropriate
$M,N\to \infty$ limit of $\pi/2\p = N/M$.  In this limit we have
 $R_D=0$ by (\ref{s2f}).  It can be verified that $R_{S1}$ and $R_{S2}$ in (\ref{s1f}) 
and (\ref{R3})  can be rewritten in 
equivalent forms
\bea
R_{S1} &=& \sum_{n=1}^{M-1} \frac 1 n \tan(n\p) \sin[2n(\p-v)] \nonumber \\
   && +\quad \sum_{n=1}^{M-1} \tan(n\p)\bigg[ \sin(2n\p) \int_{2v} ^{\pi/2n} 
\cot (Mx) \cos(nx)dx\nonumber \\
&& \hskip2cm + \cos(2n\p) \int_0^{2v} \cot (Mx) \sin(nx)dx \bigg], \nonumber \\
R_{S2} &=& 2\sum_{m=0}^{(N-3)/2} \cot \bigg[ \bigg(m+\frac 1 2\bigg) \frac {\pi x}{\p} \bigg] 
 \bigg\{ \frac {\sin[(2m+1)v\pi/\p]} {2m+1} \nonumber \\
&& \hskip2cm -\frac {M} N \int_0^{2v} \cot (Mx) \sin \bigg[\bigg(m+\frac 1 2\bigg) \frac {\pi x}{\p} \bigg]
  dx \bigg\}.\label{RR13}
\eea
In the large $M,N$ limit, the integrals in (\ref{RR13}) vanish.  Then $R_{S1}$ becomes the first term
in (\ref{noteven}), $R_{S2}$ becomes the second term in (\ref{noteven}) and
one recovers (\ref{noteven}) from (\ref{psi33}).
Similarly (\ref{psi33}) can be reached by taking an appropriate limit of (\ref{noteven})
or (\ref{noteven1}). Particularly, in the case of the points $u=0$ excluded in (\ref{Mv}), 
$\x(\p,v)$ can be 
compute by taking an appropriate large $M,N$ limit of (\ref{noteven1}).
 
\medskip
 We would like to thank Professor Shin-Nan Yang for the hospitality
at the Center for Theoretical Physics, Taipei, where this work is initiated. 
LCC is supported by a travel grant of the 
Institute of Mathematics, Academia Sinica, Taipei. FYW wishes to thank W. T. Lu for a useful
conversation.

\end{document}